\documentclass[aps,prl,twocolumn]{revtex4-2}
\usepackage{amssymb, amsmath, mathtools}
\usepackage{subfig}
\usepackage{graphicx}
\usepackage{xcolor}

\begin{document}

\title{Superluminal Propagation of Composite Collective Modes in Superconductor-Ferromagnet
Heterostructures}
\author{Pascal Derendorf, Anatoly F. Volkov, and Ilya M. Eremin}
\affiliation{Theoretische Physik III, Ruhr-Universit\"at Bochum,
  D-44780 Bochum, Germany}
\date{\today }

\begin{abstract}
Superconductor/ferromagnet/superconductor Josephson junctions are paradigmatic systems for studying the delicate interplay of superconductivity and magnetism via proximity effects as well as their composite excitations. Here, we analyse the collective modes (CM) in such a heterostructure by taking into account the interplay between the de-magnetisation field $H_{dem}$ and the field caused by the anisotropy of the ferromagnet $H_{an}$, which was previously neglected. It turns out that the spectrum of composite collective modes, $\omega(k)$, has a qualitatively different form in the case of $H_{dem}<H_{an}$ and of $H_{dem}>H_{an}$. In the first case, the 
dependence $\omega(k)$ has the same form as in previous studies, whereas in the second case, the spectrum looks completely different.  In particular, for moderate or weak anisotropy in ferromagnet the
group velocity of collective modes demonstrates inflection point where the group velocity become infinite and is superluminal. Furthermore, this point separates purely real and complex-conjugate solutions for the collective modes and is also {\it exception point}. We show that the difference of the CMs spectra can be revealed by Fiske experiment, i.\,e.\,by measuring the $I-V$ characteristics in the presence of magnetic field and voltage.
\end{abstract}

\maketitle
\date{\today }

\section{Introduction} Quantum states of matter in solids typically exhibit collective excitations like phonons, spin waves, or plasma excitations. Although these individual excitations are relatively well understood, the interaction between them once their energy spectra intersects represents an interesting field of composite excitation where the properties of two different modes interfere resulting in unique physical properties. Recent examples of this phenomenon are magnon polaron formed by selectively coupled coherent magnon and phonon modes in surface patterned ferromagnet \cite{Godejohann2020}, and skyrmion-vortex topological excitations formed in superconductor-ferromagnet heterostructures \cite{Petrovic21}. In general the energy crossing of the spectral characteristics of various collective modes (CM)
is well known in many branches of physics like in superconductors \cite%
{AnishVolkovEfetovPRB_07,BasovPRR_20,MillisPRB_20,Muller21}, ferromagnets \cite{AkhieserBook68}), in plasma \cite{VedenovUFN64,LifshitzPitaevPhysKinetics_V10}, in optics (see for example \cite{AgranovichUFN75,BasovRMP11}) etc. In particular, such a crossing of the CM spectra can occur in two-band and
single-band superconductors (see \cite{AnishVolkovEfetovPRB_07}, \cite{BasovPRR_20} and references therein). The spectra crossing in two-band
superconductors \cite{AnishVolkovEfetovPRB_07} corresponds to interception
of the hard mode branch (the Leggett mode \cite{Leggett66}) and a sound-like
branch (the Carlson-Goldman mode \cite{Carlson-GoldmanPRL75}). A detailed
analysis of CM in single-band superconductors and their detection via near
field nano optics was carried out in a recent article \cite{BasovPRR_20}.
The authors analyzed such CM and their interactions as Higgs,
Bardasis-Schrieffer, Carlson-Goldman modes, as well as plasmons.

\begin{figure}[tbp]
\includegraphics[width=0.8\columnwidth]{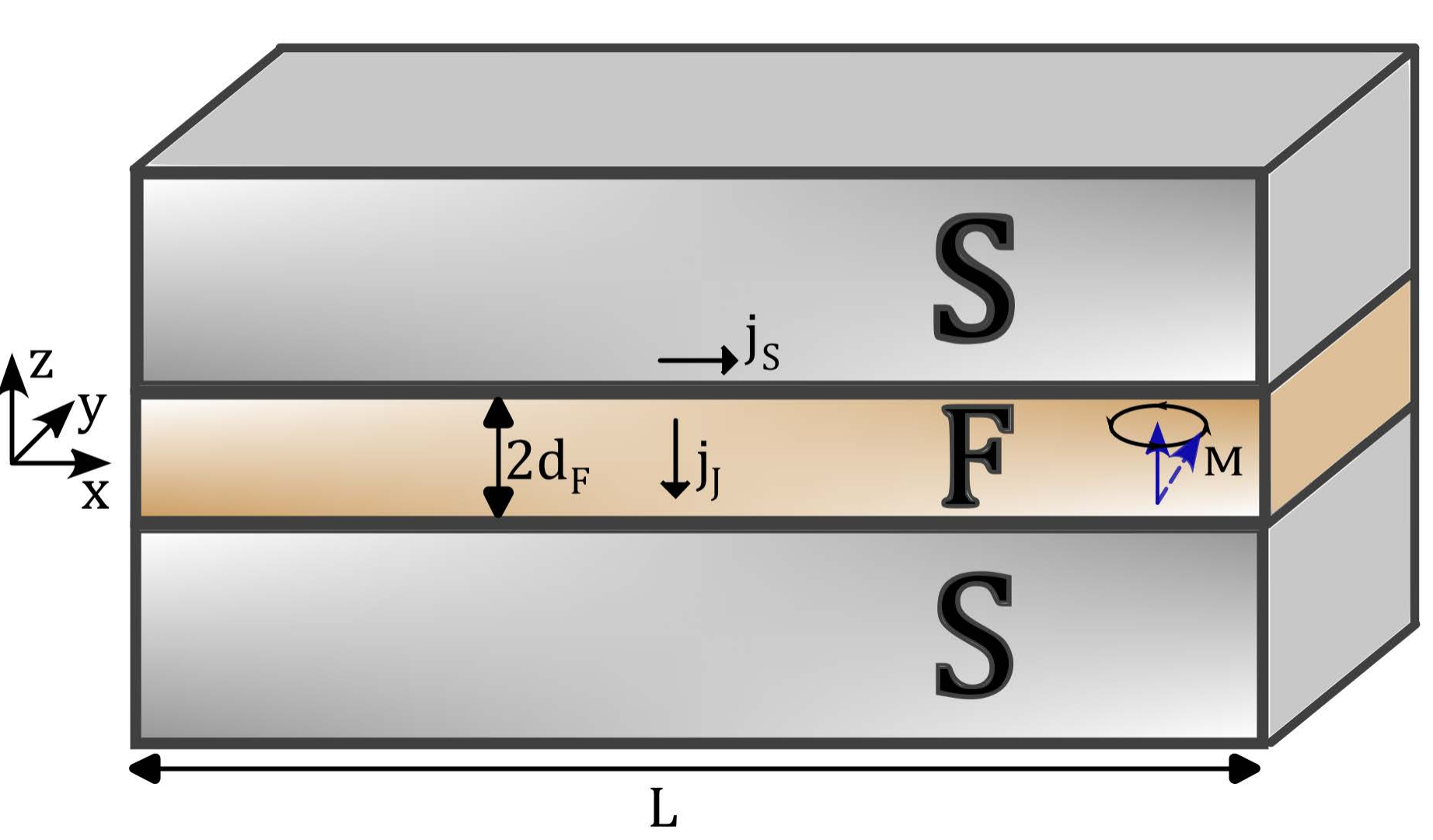}
\caption{(Color online) Schematic structure of the S/F/S heterostructure. Two superconductors (S) are connected via ferromagnet with equilibrium magnetisation, $M_{0}$, pointing along $z$-direction. Due to the time-dependent Josephson phase difference $\phi(t)$ the magnetization in ferromagnet start to precess and acquire the in-plane component; $2d_F$ is a thickness of a ferromagnet, while $j_S$ and $j_J$ refer to the supercurrent and Josephson current, respectively.}
\label{fig1}
\end{figure}

Another example of crossed branches is the magnetic S/F/S Josephson junction
(see Fig.\ref{fig1}), where the spectra of magnetic and Swihart/Josephson plasma
waves can intersect under appropriate conditions (see \cite{V-EfetovPRL09}
and subsequent papers \cite{Maekawa11,Silaev-PRB23,China23}). The crossed
spectral characteristics of CM can split even in the presence of a weak
coupling between these modes. The initial theoretical analysis was carried out based on linearized dynamic equations for the phase difference $%
\varphi$ in superconductors S and for a deviation $\mathbf{m}_{\perp}(t)$ of
the magnetisation vector $\mathbf{M}$ in the ferromagnetic film F from its
stationary orientation in the case of a time-dependent phase difference $%
\varphi(t)$. Although the models considered in \cite{V-EfetovPRL09,
Maekawa11,Silaev-PRB23} were slightly different, the modified spectrum $%
\omega(k)$ shows a characteristic avoided crossing form, characteristic for hermitian systems. Also in bulk ferromagnetic superconductors the excitation of the spin waves was also studied theoretically \cite{Braude2004,mukherjee2024}.
Experimentally, electrodynamical effects in S/F structures  were studied recently in Refs. \cite{PfeifferPRB08,RyazanUstin21,Yao2021,Ryazanov22,Ryazanov23,Borst2023,Massarotti24}.
Furthermore, magnetic superconducting heterostructure is an interesting playground to study exciting quantum phenomena such as $\pi$ -
Josephson coupling, long-range odd-frequency Cooper-pairing triplet component etc \cite{HouzetPRL08, BuzdinPRB21,Johnsen2021,Yu2022,BuzdinPRB23,SilaevCondMat22,SilaevPRL19,BobkovaPRB20,BergeretPRAppl20,Bobkova2022}, see also further reviews  \cite{GolubovRMP04,BuzdinRMP05,BVE_RMP05,EschrigRev15,LinderRob_NaturePhys15,LinderBalRMP19,BlamireAPL20,BirgeRev_APL24}. 
\textcolor{black}{Furthermore, the mutual interplay of the dynamics of the magnetic moment $\mathbf{M}$ and of the Josephson current in SFS Josephson junctions (JJ) has been also studied theoretically in Refs. \cite{Cai2010,Chudnovsky2016,Chudnovsky2017,Ghosh2017,Nashaat2018}  yet the influence of a magnetic mode (magnetic resonance) on the propagation of the collective modes has not been studied.}  

Here, we reconsider the problem of the collective modes in SFS heterostructures using Maxwell and Landau-Lifshitz equations and show that its spectrum is by far richer than anticipated previously. We show that the shape of the dispersion curve $%
\omega (k)$ near the crossing point does not depend much on 
the Josephson coupling, but crucially depends
on the strength of the magnetization anisotropy. In the case of a strong anisotropy the
dispersion curve has a typical avoided crossing form, characteristic for Hermitian systems,  similar to that obtained in Ref. \cite%
{V-EfetovPRL09, Maekawa11,Silaev-PRB23}. However, in the case of moderate or
weak anisotropy, the form $\omega (k)$ changes drastically. In particular,
the group velocity $v_{gr}=d\omega (k)/dk$ acquires two branches separated by the region of complex solutions ({\it evanescent modes}). Furthermore, at the boundary regions near the crossing points region the group velocity of the composite collective modes shows superluminal propagation, i.e. formally  
exceed the velocity of light $c$.  This fact does not mean that the causality principle is violated (see discussion of this phenomenon in optics in Refs. %
\cite{Superlumin11,SuperluminPRL11}). At the same time, this behavior does not have analog in the hermitian quantum mechanics and is characteristic for the non-hermitian systems and the inflection point separating real and complex-valued solutions are {\it exception points}. We also discuss the possibilities of detecting differences between modified spectra in experiments (Fiske experiments \cite{FiskeRMP64}) and briefly mention the effects of these
spectra on the propagation of nonlinear excitations in Josephson SFS junctions such as solitons etc \cite{KulikBook72,BaroneBook82,MalomedRMP89,Ustinov98,Newell1980,Belova1997,Kadomtsev1971,Kadomtsev1994,1986ii}.
\textcolor{black}{Interestingly enough the change in the shape of the dispersion curves from the weak to the strong anisotropy does not significantly depend on the strength of the Josephson coupling. This circumstance removes serious restrictions on the thickness of the F layer in the possible experimental realization.}

We consider the planar SFS junction as shown in Fig.\ref{fig1} and the ferromagnet is either a metal or an insulator \textcolor{black}{with anisotropy field $\mathbf{H}_{an}=4\pi KM_{0}\mathbf{n}_{z}$ directed along the $z$-axis. We assume that} in the absence of a phase difference, the magnetisation vector $\mathbf{M}_{0}$ is uniform (no domain walls) and perpendicular to the plane of the junction: $%
\mathbf{M}_{0}=\mathbf{n_{z}}M_{0}$, where $\mathbf{n_{z}}$ is a unit vector
perpendicular to the plane of the F layer. \textcolor{black}{This assumption corresponds to the experimental situation \cite{Petkovic2009} and to the previous considerations.\cite{V-EfetovPRL09, Maekawa11}}.
\textcolor{black}{Note that in general case the vector $\mathbf{M}_{0}$ can be written as $\mathbf{M}_{0}=M_{0}(\sin\theta,0,\cos\theta)$. The orientation of $\mathbf{M}_{0}$ in the stationary case is then determined by the equation $(K-\cos\theta)\sin\theta=0$ which has two solutions: (i) $\sin\theta = 0$ (out-of-plane orientation) and (ii) $\cos\theta = K$ at $\mid K \mid <1 $  and the vector $\mathbf{M}_{0}$ is tilted relative to the $z$ axis.}

If there is a time-dependent
phase difference $\varphi (t)$, the vector $\mathbf{M}$ may begin to precess
which leads to the appearance of a transverse component $\mathbf{m}_{\perp }$, i.\,e.\,the magnetisation vector acquires the form

\begin{equation}
\mathbf{M}(t)=M_{0}\mathbf{n_{z}}+\mathbf{m}_{\perp}(t)  \label{1}
\end{equation}%
where the vector $\mathbf{m}_{\perp }=(m_{x},m_{y},0)$. Therefore, the
magnetic field $\mathbf{H}$ can be represented in the form
\begin{equation}
\mathbf{H}(t)=H_{0}\mathbf{n_{z}}+\mathbf{h}_{\perp}(t)  \label{2}
\end{equation}%
The unperturbed magnetic field $H_{0}$ is equal to the demagnetisation field
$H_{0}=-4\pi M_{0}$, so that the magnetic induction in a stationary case, 
$B_{0}=H_{0}+4\pi M_{0}$, is zero both in the F film and in the superconductors S,
where $H_{0}=B_{0}=0$. 

\subsection{Perturbation of the magnetic field.} 
The perturbed {\it ac} magnetic field 
$\mathbf{h}_{\perp }$ can be found from the Maxwell equation (Faraday's law)
\begin{equation}
\mathbf{\nabla }\times \mathbf{E}=-\frac{1}{c}\partial _{t}(\mathbf{h}_{\bot }+4\pi
\mathbf{m}_{\bot })\text{,}  \label{H1}
\end{equation}
in a way similar to that given in Refs.\,\cite{BaroneBook82,Maekawa11}. We
integrate this equation in the planes ${(y,z)}$ and ${(x,z)}$ and use the
Josephson relation $2eV=\hbar\partial _{t}\varphi$ to obtain
\begin{equation}
\mathbf{h}_{\perp }=-4\pi \tilde{d}_{F}\mathbf{m}_{\perp }+a(\mathbf{n_z\times
\nabla }\varphi )\text{,}  \label{H5}
\end{equation}%
where $a=\Phi _{0}/(4\pi (\lambda _{\omega }+d_{F}))$, $\tilde{d}
_{F}=d_{F}/(\lambda _{\omega}+d_{F})$ is a dimensionless thickness of the F
film which is assumed to be a small parameter, and $\Phi _{0}=hc/2e$ is the
magnetic flux quantum. Here $\lambda _{\omega }$ is the penetration depth of
an ac magnetic field; it is defined as $\lambda_{\omega
}^{-2}=\lambda_{L}^{-2}+\delta_{sk}^{-2}$, $\lambda_{L}$ is the static
London penetration depth, and $\delta_{sk}^{2}=c^{2}/(4\pi
\sigma_{\omega}i\omega)$ is the square of the skin depth (in the case of a magnetic insulator F this length is infinite). 

\subsection{Variation of magnetisation $\mathbf{m}_{\perp }$.} 
The dynamics of the magnetic moment $\mathbf{M}$ is described by the Landau-Lifshits
equation\cite{LL,Landau,KittelBook67,MartinBook67,AkhieserBook68,LifshitzPitaevStatPhys2-77}, which includes the Gilbert damping term
\begin{equation}
\frac{d\mathbf{M}}{dt}=\gamma \mathbf{M}\times \mathbf{B}+\frac{\alpha
}{M_{0}}\mathbf{M\times }\frac{d\mathbf{M}}{dt}.  \label{9}
\end{equation}
Here, $\mathbf{B}=\mathbf{H}_{dem}+\mathbf{H}_{an}+\mathbf{h}_{\perp
}+4\pi \mathbf{M}$, and $\mathbf{H}_{dem}$ is a demagnetisation field, which for a thin film equals $\mathbf{H}_{dem}=-4\pi M_{0}\mathbf{n}
_{z}$ in
the considered geometry, $\mathbf{H}_{an}=4\pi KM_{0}\mathbf{n}_{z}$ is an effective
field related to the uniaxial anisotropy (note the difference $4\pi$ in the  conventional definition\cite{LifshitzPitaevStatPhys2-77}) and $\mathbf{h}_{\perp}=(h_{x},h_{y},0)$. The coefficient $\gamma $ is related to the gyromagnetic
ratio $g$: $\gamma =-g\mu _{B}/\hbar $. The Gilbert damping is given by the
last term in Eq.\,(\ref{9}). Using Eqs.\,(\ref{1},\ref{2}) and linearising Eq.\,(%
\ref{9}), we arrive at
\begin{equation}
\frac{d\mathbf{m}_{\perp}}{dt}=\Omega _{m}\mathbf{n}_{z}\times \lbrack
\mathbf{h}+4\pi \kappa \mathbf{m}]_{\perp }+\frac{\alpha }{M_{0}}\mathbf{%
n_{z}\times }\frac{d\mathbf{m}_{\perp }}{dt}  \label{10}
\end{equation}
where $\Omega _{m}=\gamma M_{0}$, $\kappa=\tilde{\lambda}_{\omega}-K\approx 1-K$ and  $\tilde{\lambda}_{\omega }=\lambda _{\omega }/(\lambda _{\omega}+d_{F})\approx 1$. For simplicity we neglect  spatial dispersion in the spectrum of spin waves $\Omega _{M,k}=\Omega
_{m}[1+(kl_{m})^{2}]$.
because $kl_{m}\sim l_{m}/l_{ch}\ll 1$, where $l_{ch}$= \text{min}($l_{J},\bar{c}/\Omega_{m}$) is a characteristic length of the problem and $\bar{c}=c/\sqrt{\epsilon\lambda/d_{F}}$ is the velocity of Swihart waves \cite{Swihart61}. The magnetic length $l_{m}$ is of the order of the thickness of the domain wall. 
The field $\mathbf{h}_{\perp }$ is related to the phase difference $\varphi $ and $\mathbf{m}_{\perp }$ in Eq.\,(\ref{H5}).

In order to find a relation between $\mathbf{m}_{\perp }$ and $\varphi$ from
Eq.\,(\ref{10}), we use Eq.\,(\ref{H5}) and obtain
\begin{eqnarray}
i\omega \mathbf{m}_{\perp }-\tilde{\Omega}_{m}(\mathbf{n}_{z}\times \mathbf{%
\mathbf{m}_{\perp })} &=&-a\Omega _{m}\mathbf{\nabla }_{\bot }\varphi ,
\label{11a} \\
i\omega (\mathbf{n}_{z}\times \mathbf{\mathbf{m}_{\perp })}+\tilde{\Omega}%
_{m}\mathbf{m}_{\perp } &=&-a\Omega _{m}(\mathbf{n}_{z}\times \mathbf{\nabla
}_{\bot }\varphi )  \label{11b}
\end{eqnarray}
where $\tilde{\Omega}_{m}=4\pi \Omega _{m}\kappa$. 
\begin{equation}
 \tilde{\Omega}_{m}=4\pi \Omega _{m}\kappa = 4\pi \Omega _{m}(1-K)
\text{, }  \label{12a}
\end{equation}%
\textcolor{black}{In the case of a tilted vector $\mathbf{M}_{0}$, the parameter $\kappa$ is replaced by $\kappa_{\theta}=\cos\theta - K -\sin^2\theta$}.

Note that  $\tilde{\Omega}_{m}$ can be both
positive and negative since the anisotropy constant $K$ varies in a wide interval \cite{LifshitzPitaevStatPhys2-77} . For the $K<1 $, i.\,e.\, moderate
anisotropy it is positive. At present we neglect the Gilbert damping (a
detailed analysis of the influence of the Gilbert damping on the behaviour of
the so-called $\phi_{0}$ SFS Josephson junctions was studied for example in Ref. \cite%
{ShukrinovPRB21}). We find for $\mathbf{m}_{\perp }$
\begin{equation}
\mathbf{m}_{\perp }=a\frac{\Omega _{m}}{\mathcal{D}_{m}}[i\omega \mathbf{%
\mathbf{\nabla }_{\bot }\varphi }+\tilde{\Omega}_{m}(\mathbf{n_{z}}\times
\mathbf{\mathbf{\nabla }_{\bot }\varphi )]}\text{, }  \label{12}
\end{equation}%
where $\mathcal{D}_{m}=\omega ^{2}-\tilde{\Omega}_{m}^{2}$. Thus, Eq.\,(\ref%
{H5}) for the magnetic field $\mathbf{h}_{\perp}$ can be rewritten as
\begin{equation}
\begin{split}
\mathbf{h}_{\perp}=&a\{(\mathbf{n_{z}\times \nabla }\varphi )[1-4\pi\tilde{d}_{F}%
\frac{\tilde{\Omega}_{m} \Omega_{m}}{\mathcal{D}_{m}}]\\&-4\pi\tilde{d}_{F}%
\frac{\Omega _{m}i\omega }{\mathcal{D}_{m}}\mathbf{\mathbf{\nabla }_{\bot
}\varphi \}}\text{,}  \label{13}
\end{split}
\end{equation}
where we neglected the terms of the order of $\tilde{d}_{F}^{2}$. 

\subsection{Spectrum of Collective Modes.}
In the derivation of a dynamic equation for the
phase difference, we follow Refs.\,\cite{KulikBook72,BaroneBook82}. Consider
the $z$-component of the Maxwell equation inside the F layer
\begin{figure}[htbp]
\centering
    \includegraphics[width=0.47 \textwidth]{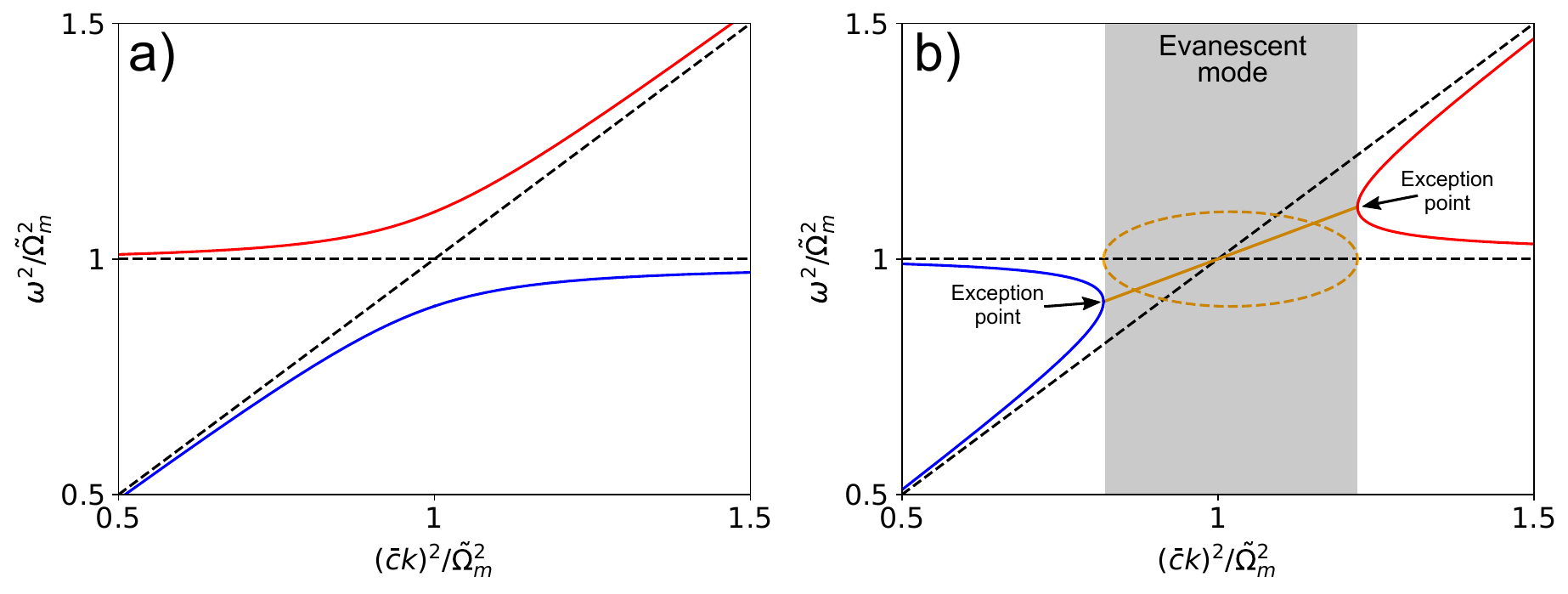}
\caption{(Color online.) The spectrum of CM calculated using Eq.(\ref{C4a}) and Eq.(\ref{C4b}), respectively (see also Eqs.(\ref{C1},\ref{C2})). We take the parameter $4\Tilde{d}_F R=0.2$. The dotted lines refer to the non-interacting dispersions. In (b) the grey area refer to the imaginary solutions where the orange dashed line represents the imaginary part of the dispersion, which was shifted to $\tilde{\Omega}_{m}^2$ for clarity. The points connecting purely real and complex conjugate solutions are {\it exception} points, which correspond formally to the infinite group velocity, i.e. superluminal propagation.}
\label{fig2}
\end{figure}
\begin{equation}
(\nabla _{\bot }\times \mathbf{h}_{\bot })_{z}=\frac{4\pi }{c}(j_{c}\sin
\varphi +\sigma _{\omega }E_{z})+\frac{\epsilon }{c}\frac{\partial E_{z}}{%
\partial t}\text{,}  \label{14}
\end{equation}%
Here the last term is the displacement current in the magnetic insulator F
and $\sigma _{\omega }E_{z}$ is the quasiparticle current. In the case of a magnetic insulator F and low temperatures, this term can be considered small. Note also that we consider the magnetic field $\mathbf{h}_{\bot }$ almost
$z$-independent in the F layer because a characteristic length over which
the function $\mathbf{h}_{\bot }(\mathbf{r}_{\bot },z)$ changes is much
larger than $d_{F}$. Combining Eqs.\,(\ref{13}-\ref{14}) and assuming that the
parameter $\tilde{d}_{F}\ll 1$ is small, we arrive at the equation for the
phase difference
\begin{equation}
\frac{1}{\bar{c}^{2}}\left(\frac{\partial ^{2}\varphi}{\partial t^{2}}+\gamma _{J}%
\frac{\partial \varphi }{\partial t}\right)-\nabla _{\bot }^{2}\varphi
+l_{J}^{-2}\sin \varphi =-\tilde{d}_{F}R\frac{4\pi \tilde{\Omega} _{m}^{2}}{%
\mathcal{D}_{m}}\nabla _{\bot }^{2}\varphi  \label{16}
\end{equation}%
where $l_{J}^{2}=ac/(4\pi j_{c})=(c/4\pi j_{c})(\Phi _{0}/4\pi \lambda _{L})$, $\bar{c}=c/\sqrt{\epsilon\lambda/d_{F}}$ is the velocity of the Swihart waves and $\gamma
_{J}=4\pi \sigma _{\omega }$. As it is noted above, the coefficient $R=%
{\Omega}_{m}/\tilde{\Omega} _{m}$ may be both positive (weak or moderate
anisotropy) or negative (strong anisotropy).

The spectrum of collective modes can be easily found by linearizing Eq.(\ref{16}) and substituting $\varphi (r_{\bot },t)\sim \varphi (0,0)\exp (i\omega t-i\mathbf{k}_{\bot}\cdot 
\mathbf{r}_{\bot })$. Then, this equation can be written in the form
\begin{equation}
\lbrack \omega ^{2}-(\Omega _{J}^{2}+Q^{2})][\omega ^{2}-\tilde{\Omega}%
_{m}^{2}]=-\tilde{d}_{F}R Q^{2}\tilde{\Omega}_{m}^2,  \label{C1}
\end{equation}
where $Q=\bar{c}k_{\perp}$.
For the sake of clarity we also introduce the variables $X=Q^{2}$, $Y=\omega ^{2}$, $Y_{J}=\Omega _{J}^{2}$
, $Y_{m}=\tilde{\Omega}_{m}^{2}$, and can write this equation as follows
\begin{equation}
    \left[Y-(X+Y_{J})\right]\left[Y-Y_{m}\right]=-\Tilde{d}_{F}RXY_{m}. \label{C2}
\end{equation}
The right hand side of this equation contains a small parameter $\tilde{d}_{F}$, responsible for the coupling of magnetic and Swihart/Josephson modes. If $\tilde{d}_{F}=0$ they are decoupled and one gets $Y_{0} = Y_{m}$, $X_{0} = Y_{m}-Y_{J}\geqslant 0$. For simplicity, we also neglected damping in both modes and the dispersion of the magnetic mode which is small compared to the dispersion of Swihart modes at not too large wave vectors $k$. Due to the coupling, the spectrum
is modified and the spectrum shows two types of the avoided crossing point disappears (see Fig.\ref{fig2}). The small
deviation from $(X_{0},Y_{0})$ are defined as $y=Y-Y_{0}$, $x=X-X_{0}$%
. The dispersion curve for CM near the crossing point can be easily found
from Eq.\,(\ref{C2}). In particular, 
for a strong anisotropy, i.\,e.\,for $K>1 $, the dispersion curve can be
written as
\begin{equation}
y_{1,2}=\frac{1}{2}[x\pm (x^{2}+4\tilde{d}_{F}|R|Y_{m}(X_{0}+x))^{1/2}].
\label{C4a}
\end{equation}%
and is qualitatively similar to the spectra obtained in Refs.\,\cite%
{V-EfetovPRL09,Maekawa11,Silaev-PRB23} as shown in Fig.\,\ref{fig2}(a).

However, in the case of positive $R$ ($K<1$), the form of the spectra near the crossing point has the form 
\begin{equation}
y_{1,2}=\frac{1}{2}[x\pm (x^{2}-4\tilde{d}_{F}RY_{m}(X_{0}+x))^{1/2}].
\label{C4b}
\end{equation}%
and is shown in Fig.\,\ref{fig2}(b) with very unusual avoided crossing behavior. 
In particular, there is a wavevector region where the solution is imaginary and, moreover, near the inflection point the group velocity of the composite excitations formally tends to infinity and exceeds the speed of light (inflection point).   
Taking into account damping would eliminate this singularity, but the maximum
group velocity in this case depends on the damping rate.
Away from the inflection point the group velocity becomes negative, indicating potential instability of this homogeneous solution. Interestingly, the $k^2$ region of the imaginary solutions might be related to the scale of spatial inhomogeneity in the system. \textcolor{black}{In particular,   
for positive $R$ the perturbations with some wave vectors $ k$ in the region $|x|<2Y_{m}\sqrt{\tilde{d}_{F}|R|}$ grow in time. This means an instability of an initial state.}  Observe also that our consideration is classical yet the solution, represented in Fig.\ref{fig2}(b) would correspond to the non-hermitian case and the point separating imaginary and real modes is typically called {\it exception} point \cite{Ashida2020}.

\textcolor{black}{Let us also estimate the characteristic length of the problem and whether the regime with superluminal velocities of the composite modes is experimentally accessible. For example, if one takes $\bar{c} \approx (0.1 - 0.15) \times 10^{10}cm/sec$ and $\Omega_{M} \approx 10 GHz$, we obtain for $l_{ch} \approx \bar{c}/\Omega_{M} \approx 100 - 150 \mu m$. This is experimentally achievable quantity given the reported diameter, $D \approx 300 \mu m$ \cite{Davidson1985} and $D \approx 500 \mu m$\cite{Krasnov1997} of circular Josephson junctions in the experiments.}

\subsection{Fiske experiment.} 
Formally, group velocities  exceeding the speed of light are not forbidden and have been recently discussed in the context of  superluminal propagation in active structures\cite{Wang2000,Stenner2003,Superlumin11,SuperluminPRL11,Hrabar2013,Niang2017,Duggan2022} and the consensus is that it does not violate the causality as the signal would still propagate with velocities, smaller than the speed of light.
Nonetheless, the first question is how to
experimentally determine which spectra can be observed, i.\,e.\,whether $R$
is positive or negative. Fiske and collaborators observed
changes in the I-V curve in a tunnel SIS Josephson junction in the presence
of a weak magnetic field $H$ (the so-called Fiske steps \cite{FiskeRMP64}).
These changes are caused by the interaction of a travelling mode $\sin
\varphi _{0} = \sin (\omega _{V}t-k_{H}x)$ with a Josephson plasma mode, where $\omega _{V}=2eV/\hbar$, and $k_{H}=4\pi\lambda/\Phi_{0}$.
The theory of this effect was given in Refs.\,\cite{EckPRL64,Kulik65} and
is well understood \cite{KulikBook72,BaroneBook82}. To follow this idea 
we write the
phase difference $\varphi $ in the form: $\varphi =\varphi _{0}+\psi $,
where $\varphi _{0}=\omega _{V}t-k_{H}x$ and $\psi $ is considered to be small. It
obeys the equation $\mid $%
\begin{equation}
-\frac{1}{\bar{c}^{2}}[\partial _{tt}^{2}+\gamma \partial _{t}]\psi
+P_\omega\partial _{xx}^{2}\psi =l_{\omega J}^{-2}\sin (\omega _{V}t-k_{H}x)\text{,}
\label{1a}
\end{equation}%
We look for the solution of $\psi $-function in the form $\psi (t,x)=\text{Im}\Psi (t,x)$,
where $\Psi (t,x)=\Psi (x)\exp (i\omega _{V}t)$ and $\Psi (x)$ satisfies the
equation
\begin{equation}
\partial _{xx}^{2}\Psi (x)+k_{V\omega}^{2}\Psi (x)=k_{J\omega}^{2}\exp
(-ik_{H}x)\text{,}  \label{2a}
\end{equation}%
where $k_{V\omega}^{2}=(\omega^{2}_{V}+i\gamma \omega_{V})/(\bar{c}^{2}P_{\omega })$, $k_{J\omega}^{2}=k_{J}^{2}/P_{\omega }$ with $P_{\omega }=(\mathcal{D}_{m}-\tilde{d}_{F}R(4\pi \tilde{\Omega}_{m} )^{2})/\mathcal{D}_{m}$. Eq.\,(\ref{2a}) is complemented
with the boundary conditions: $\mathbf{h}_{\bot }(\pm \frac{L}{2})=0$ which is
equivalent to $\partial _{x}\Psi (\pm \frac{L}{2})=0$. The solution can be represented
in the form of a sum
\begin{equation}
\Psi (x)=\sum_{n=0}^{\infty }a_{n}\cos (k _{n}x)\text{,}  \label{3a}
\end{equation}%
where $a_{n}=(1/L)\int_{-L/2}^{L/2}\cos (k _{n}x)\Psi (x)dx \equiv \langle \Psi (x)\cos (k_{n}x)\rangle _{x}$ and $k
_{n}=2\pi n/L$. From Eqs.\,(\ref{2a}, \ref{3a}), we can easily obtain
\begin{equation}
a_{n}=\frac{k_{J\omega }^{2}}{k_{V\omega }^{2}-k_{n}^{2}}C_{n}\text{,}
\label{4}
\end{equation}%
where $C_{n}=\langle \exp (-ik_{H}x)\cos (k_{n}x)\rangle _{x}.$

\begin{figure}[htbp]
\centering
\includegraphics[width=0.47 \textwidth]{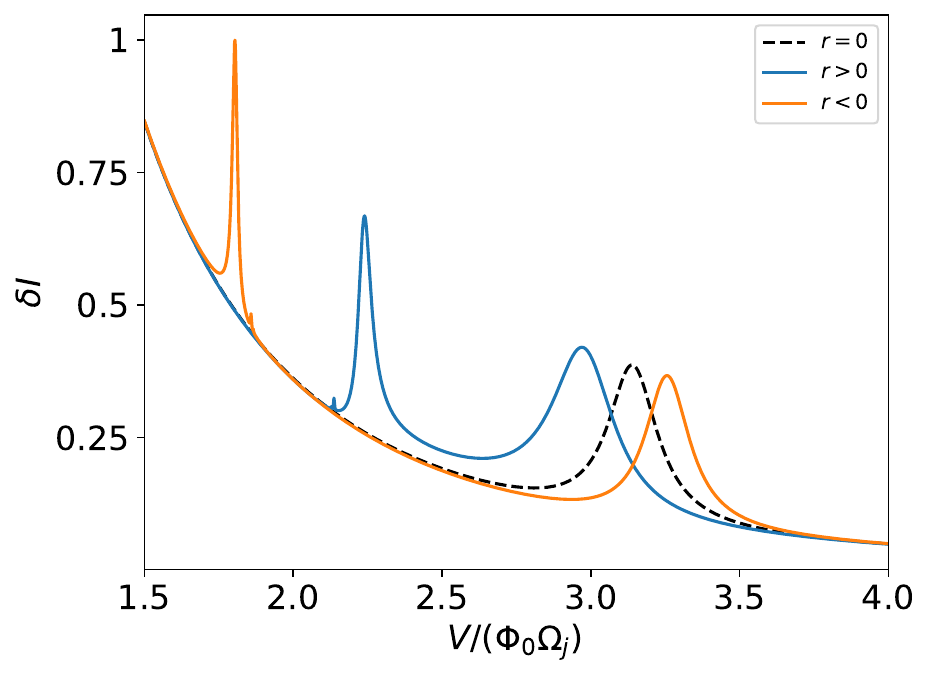}
\caption{(Color online.) The normalized (in regard to the first peak) corrections $\delta I$ to the I-V curve due to F layer as a function of $V/\Phi \Omega_j$. 
%The panel (a) refers to the situation with no damping for $D_m=\omega_v^2-\tilde{\Omega}_{m}^2$, while in the panel (b) the damping is included, i.e. $D_m=\omega_v(\omega_v-i\gamma_m)-\tilde{\Omega}_{m}^2$. 
Here, $\gamma = 0.2\Omega_j$, $r=\tilde{d}_{F}R$ and $\gamma_m = 0.1\Omega_j$ with  $\tilde{\Omega}_m=2\Omega_j$, $L=4l_j$ and $\Omega_j=1$. The dashed curve refer to the situation when the collective modes in a ferromagnet and superconductor do not interact, $r=0$ (SIS junction).}
\label{fig3}
\end{figure}

The correction to the current $\delta I_{dc}=\langle \psi (t,x)\cos (\omega
_{V}t-k_{H}x)\rangle _{x,t}$ is equal to
\begin{equation}
\delta I=\sum_{n=0}|C_{n}|^{2}\text{Im}\frac{k_{J\omega }^{2}}{(k_{V\omega
}^{2}-k_{n}^{2})}\text{.}  \label{5}
\end{equation}%
where $\theta _{H\pm }=(k_{H}\pm k_{n})L$ and $\theta _{H}=k_{H}L$.The
coefficient $|C_{n}|^{2}$ is equal to
\begin{equation}
|C_{n}|^{2}=4\left(\frac{\theta _{H}}{\theta _{H}+\theta _{n}}\right)^{2}\left[\frac{\sin
(\theta _{H}/2)}{{\theta _{H}-\theta _{n}}}\right]^{2}\text{,}
\label{5a}
\end{equation}
We plot the $\delta I_{dc}(V/\Phi_0 \Omega_j)$ in Fig.\,\ref{fig3} for positive and negative $R$, i.\,e.\,for a moderate or a strong anisotropy where we introduced $r = \tilde{d}_F R$.
If the collective modes in a ferromagnet and a superconductor do not interact ($r=0$) the correction $\delta I$ shows only one resonance as shown in Fig.\ref{fig3} by the dotted line, which reflects the position of the Swihart mode. Once $r\neq 0$, the position of the resonance (first Fiske steps in the I-V correction) shifts either to the lower value for $r>0$ (see Fig.\ref{fig2}(a) for a strong magnetic anisotropy) or to the higher value for $r<0$ (see Fig.\ref{fig2}(b) for a weak magnetic anisotropy) depending where the position of the collective mode shifts due to the coupling. This will provide information on the sign of the coupling. In addition, one can see two peaks at lower voltages V, which are absent if the F layer is replaced by a non-magnetic insulator. These peaks are caused by  excitations of a magnetic resonance in F; their positions and amplitudes depend on $r$ and on the strength of the damping, respectively.

\section{Discussion.} 
We investigated the coupling of the collective modes in the SFS heterostructure, where the branches of the magnetic spectrum and the spectrum of the Swihart/Josephson's modes may interact with each other. A
coupling of these modes leads to their splitting and the shape of the spectrum of the modified modes appears to depend crucially on the magnetic anisotropy of the ferromagnet. In the case of a strong anisotropy, the form of the modified spectra is similar to those obtained earlier \cite{V-EfetovPRL09, Maekawa11,Silaev-PRB23}. 
The most interesting case corresponds to a moderate or weak anisotropy, \textcolor{black}{ not considered previously}, when the parameter $R$ is positive. In this case, the group velocity of the coupled magnetic and
plasma modes has an anomaly in a vicinity of the crossing point. It can
exceed the speed of light (superluminal propagation) and may be both
positive and negative. Furthermore, this point separates the region in {\bf k}-space of real and imaginary solutions making it {\it exception} points. The formation of these points is a characteristic feature of non-hermitian systems. We further argue that the sign of the difference between the strengths of the magnetic anisotropy and of the demagnetization field can be probed in Fiske experiment (steps). Thus, one can determine the conditions that ensure the superluminal propagation of CM.
\textcolor{black}{In addition, in the case of weak anisotropy, the $\omega(k)$ dependence contains  both  real and imaginary parts. This means an increase in the initial perturbation in time and the transition of a system from out-of-plane orientation of {\bf M} to a more stable in-plane orientation. Such a transition can be also studied with the help of modern ultrafast pump-probe techniques\cite{devecchi2024}, where fast switching of the magnetic field has been developed.}

Note, we considered the propagation of CM of a small amplitude. It
is also of interest to study the propagation of nonlinear waves or objects
like fluxons in Josephson junctions or domain walls in the ferromagnetic
layer of the considered structure. Although studying solitons is a well investigated subject (see for example, \cite{Newell1980,Belova1997, Kadomtsev1971, Kadomtsev1994, 1986ii} and referencies therein), to the best of our knowledge, the propagation of nonlinear waves in a system with coupled CM in a system with coupled CM has not yet been investigated.
For example, the l-h-s of Eq.\,(\ref{16}) describes a moving soliton
(fluxon) if the r-h-s can be considered as a small perturbation. Then, the
velocity of the moving fluxon will slow down or accelerate depending on the
sign of the parameter $R$. This can be shown if the r-h-s is small. However,
if the parameter $\tilde{d}_{F}R(4\pi \Omega _{m})^{2}/ \mathcal{D}_{m}$ is
not small (this is the case when a characteristic frequency approaches to $%
\Omega_{m}$), a nonperturbative analysis is required. This issue as well as
the physics of superluminal propagation of the magnetoplasma waves deserves a
separate investigation.

{\it Acknowledgements.} The work is supported by the Mercator Research Center Ruhr (MERCUR) project ”Composite
collective excitations in correlated quantum materials” of
the University Alliance Ruhr.

\bibliography{literature}
\end{document}